\documentclass[journal,10pt,final]{IEEEtran}
\IEEEoverridecommandlockouts

\usepackage{cite}
\usepackage{indentfirst}
\usepackage{graphicx}
\usepackage{changepage}
\usepackage{stfloats}
\usepackage{amsfonts,amssymb}
\usepackage{bm}
\usepackage{algorithm}
\usepackage{algorithmic}
\usepackage{amsmath}
\usepackage{amsmath,amssymb,amsfonts}
\usepackage{times}
\usepackage{url}
\usepackage{textcomp}
\usepackage{xcolor}
\usepackage{color}
\usepackage{setspace}
\usepackage{enumitem}
\usepackage{float}
\usepackage{diagbox}
\usepackage[T1]{fontenc}
\usepackage[utf8]{inputenc}
\usepackage{authblk}
\usepackage{threeparttable}
\usepackage{booktabs}
\usepackage{footnote}
\usepackage{graphicx}
\usepackage{pifont}
\usepackage{subfigure}
\usepackage{mathrsfs}

\begin{document}

\title{Energy-Aware UAV-Enabled Target Tracking: Online Optimization with Location Constraints}

\author{\IEEEauthorblockN{Yifan~Jiang, Qingqing~Wu, \textit{Senior Member, IEEE}, Wen~Chen, \textit{Senior Member, IEEE}, Hongxun Hui, \textit{Member, IEEE}}
\thanks{Yifan~Jiang is with the State Key Laboratory of Internet of Things for Smart City, University of Macau, Macao 999078, China (email: yc27495@umac.mo) and also with Shanghai Jiao Tong University, Shanghai 200240, China. 
Qingqing~Wu and Wen~Chen are with the Department of Electronic Engineering, Shanghai Jiao Tong University, Shanghai 200240, China (e-mail: \{qingqingwu@sjtu.edu.cn; wenchen@sjtu.edu.cn\}).
Hongxun~Hui is with the State Key Laboratory of
Internet of Things for Smart City and Department of Electrical and Computer
Engineering, University of Macau, Macao, 999078 China (email: hongxunhui@um.edu.mo).}}

\maketitle

\begin{abstract}
    For unmanned aerial vehicle (UAV) trajectory design, the total propulsion energy consumption and initial-final location constraints are practical factors to consider. 
    However, unlike traditional offline designs, these two constraints are non-trivial to concurrently satisfy in online UAV trajectory designs for real-time target tracking, due to the undetermined information.
    To address this issue, we propose a novel online UAV trajectory optimization approach for the weighted sum-predicted posterior Cramér-Rao bound (PCRB) minimization, which guarantees the feasibility of satisfying the two mentioned constraints.
    Specifically, our approach designs the UAV trajectory by solving two subproblems: the candidate trajectory optimization problem and the energy-aware backup trajectory optimization problem. 
    Then, an efficient solution to the candidate trajectory optimization problem is proposed based on Dinkelbach's transform and the Lasserre hierarchy, which achieves the global optimal solution under a given sufficient condition. 
    The energy-aware backup trajectory optimization problem is solved by the successive convex approximation method. 
    Numerical results show that our proposed UAV trajectory optimization approach significantly outperforms the benchmark regarding sensing performance and energy utilization flexibility. 
\end{abstract}



\section{Introduction}

Unmanned aerial vehicle (UAV) enabled wireless sensing has recently received growing research interest due to emerging applications such as integrated sensing and communication (ISAC) \cite{KTMeng2024WC-UAV,qqw2021JSAC,qqw2024-PIEEE-IS,XinyiWang-UAVmagazine}. 
In addition to providing wireless channels with high line-of-sight (LoS) probability, the controllable maneuverability of UAVs provides additional design degree-of-freedoms (DoFs) for sensing performance enhancement, such as trajectory optimization \cite{KTMeng-2023-TWC-UAV-IPSAC,KTMeng-2022-WCL-UAV-IPSAC}.
However, unlike airborne radar systems, UAV mobility is constrained by size, weight, and power constraints, especially the limited onboard energy \cite{qqw2021JSAC,YZ2019PIEEE,ZY-EM-Rotary-2019-TWC}. 
Therefore, energy-aware UAV trajectory optimization is one of the most critical design issues in emerging UAV-enabled sensing or ISAC systems.

Prior research on energy-aware trajectory design in UAV-enabled sensing systems can be generally classified into two categories: offline designs \cite{Khalili2023GLBCOM,ShuyanHu2022TCOM} and online designs \cite{XiaoyeJing2024TWC,JunWu-2023-IOT-UAVISAC}. 
Particularly, in offline designs, the waypoints of a complete UAV trajectory are jointly optimized. 
However, offline designs are not applicable in most dynamic scenarios, such as target tracking, where the UAV trajectory is required to be dynamically adjusted according to the target movement or environmental changes. 
Comparatively, online designs can respond to such changes timely because the UAV trajectory is sequentially optimized based on real-time measurement of the target motion state. 
For example, a multi-stage UAV trajectory optimization approach was proposed in \cite{XiaoyeJing2024TWC}, where a part of the UAV trajectory was optimized within each stage for maximizing an ISAC trade-off objective function based on the previously measured results, subject to a total propulsion energy consumption constraint.
In \cite{JunWu-2023-IOT-UAVISAC}, the UAV trajectory within each time slot was optimized to minimize the predicted posterior Cramér-Rao bound (PCRB) for the estimated target motion state. 
Nonetheless, the UAV trajectory optimization algorithms proposed in \cite{XiaoyeJing2024TWC} and \cite{JunWu-2023-IOT-UAVISAC} only generally lead to suboptimal solutions. 
Furthermore, due to the lack of controlling the complete trajectory, the approaches proposed in \cite{XiaoyeJing2024TWC} and \cite{JunWu-2023-IOT-UAVISAC} are infeasible when considering additional initial-final location constraints, which model practical scenarios launching and landing in predetermined locations \cite{YZ2019PIEEE}. 

Motivated by the above issues, we study the online UAV trajectory design in a UAV-enabled target tracking system, where a UAV tracks a moving target via the classic extended Kalman filtering (EKF) method. 
Specifically, the UAV trajectory within a short period is optimized to minimize the weighted sum of predicted PCRBs for the estimated target motion state of the next period. 
Meanwhile, the complete UAV trajectory is required to meet the total propulsion energy consumption and initial-final location constraints. 
The main contributions of this work are summarized as follows: 
\textbf{1)} We propose a novel UAV trajectory optimization approach, which solves two subproblems to optimize a candidate UAV trajectory for the weighted sum-predicted PCRB minimization and a backup UAV trajectory for the propulsion energy minimization, respectively. 
Our proposed approach achieves flexible energy utilization for sensing performance maximization and ensures the feasibility of the optimized UAV trajectory. 
\textbf{2)} We propose efficient solutions to the subproblems based on Dinkelbach's transform, the Lasserre hierarchy, and the successive convex approximation (SCA) method. 
The proposed solution to the candidate UAV trajectory optimization problem is globally optimal under a provided sufficient condition. 
\textbf{3)} Simulation results validate our proposed approach's effectiveness and illustrate a significant sensing performance improvement and a more flexible energy utilization over the benchmark.

\section{System Model and Problem Formulation}\label{SecII}

\begin{figure}[!t]
    \centering
    \includegraphics[width=0.47\textwidth]{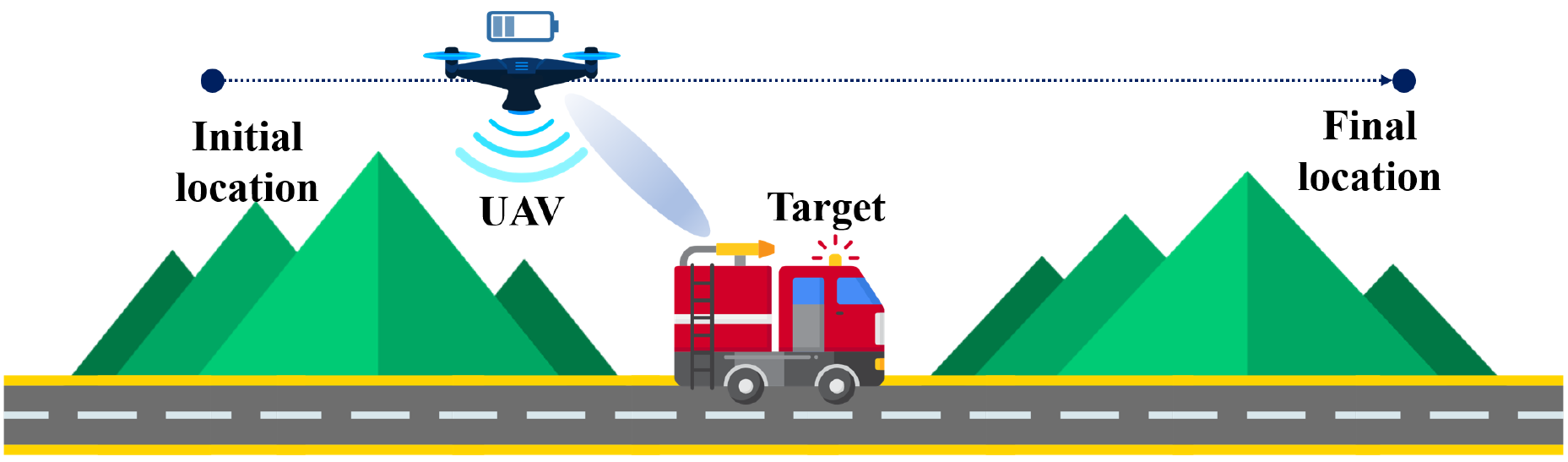}
    \vspace{-1mm}
    \caption{A UAV-enabled target tracking system with initial and final locations.}
    \label{fig::sm}
    \vspace{-4mm}
\end{figure}

As shown in Fig.\ref{fig::sm}, we consider a UAV-enabled target tracking system, where a UAV is dispatched from the initial location $x_{\text{I}}$ to the final location $x_{\text{F}}$ within a predetermined flight duration $T$ in order to track a moving ground target.
To be specific, it is assumed that both the UAV and the target motion state, i.e., their position and velocity, remain constant during a short period $\Delta T$ \cite{KTMeng-2023-TWC-UAV-IPSAC}.  
As such, by discretizing the whole flight duration $T$ into $N=T/\Delta T$ time slots, the relative motion state of the target with respect to the UAV can be expressed as $\mathbf{x}_{n} = [x_{n}, v_{n}]^{T}$, where $x_{n}$ and $v_{n}$ denotes the relative position and velocity, respectively. 
As an initial work, it is assumed that both the UAV and the target follow the one-dimensional horizontal movement.
In addition, the UAV flies at a fixed altitude $H$ and is equipped with a uniform linear array (ULA) with $N_{\text{t}}$ transmit antennas and a ULA with $N_{\text{r}}$ receive antennas parallel to the horizontal path.

\subsection{Signal and Measurement Model}

Let $s_{n}(t)\in\mathbb{C}$ denote the equivalent baseband signal transmitted with the power $P_{\text{A}}$ at the $n$th time slot. 
Due to the LoS-dominant air-to-ground channel, the large-scale sensing channel gain can be represented by $G_{n} = \beta_{\text{r}} d_{n}^{-4}$ with $\beta_{\text{r}} = \lambda^{2}\varepsilon/(64\pi^{3})$, following the free-space path-loss model \cite{YFJ2024CL}, where $\lambda$ denotes the carrier wavelength, $\varepsilon$ denotes the target radar cross section and $d_{n}$ denotes the distance from the target to the UAV at the $n$th time slot. 
Then, the sensing channel can be expressed as $\mathbf{h}_{n} = \sqrt{G_{n}} e^{-j4\pi\frac{d_{n}}{\lambda}}\mathbf{b}(\phi_{n})\mathbf{a}^{H}(\phi_{n})$, where $\mathbf{a}(\phi) = [e^{j\frac{\pi(N_{\text{t}}-1)\cos{\phi}}{2}}, ..., e^{-j\frac{\pi(N_{\text{t}}-1)\cos{\phi}}{2}}]^{T}$ and $\mathbf{b}(\phi) = [e^{j\frac{\pi(N_{\text{r}}-1)\cos{\phi}}{2}}, ..., e^{-j\frac{\pi(N_{\text{r}}-1)\cos{\phi}}{2}}]^{T}$ denote the UAV transmit and receive array response vectors, respectively, and $\phi_{n}$ denotes the elevation angle of the geographical path connecting the UAV to the target at the $n$th time slot. 
The echo signals can be represented by $\mathbf{r}_{n}(t) = \sqrt{P_{\text{A}}}\mathbf{h}_{n}e^{j2\pi\mu_{n}t}\mathbf{f}_{n}s_{n}(t-\tau_{n}) + \mathbf{z}_{\text{r},n}(t)$, where $\mu_{n}$ denotes the Doppler shift, $\mathbf{f}_{n}$ denotes the transmit beamforming vector, $\tau_{n}$ denotes the round-trip time delay, and $\mathbf{z}_{\text{r},n}(t)\in\mathbb{C}^{N_{\text{r}}\times 1}$ denotes the complex additive white Gaussian noise with zero mean and variance of $\sigma^{2}$.

At the $n$th time slot, the elevation angle, the distance, and the Doppler shift can be measured via the maximum likelihood estimation and the matched filtering, respectively \cite{YFJ2024CL}. 
To be specific, the measured results can be given by $\hat{\phi}_{n} = \phi_{n} + z_{1,n}$, $\hat{d}_{n} = d_{n} + z_{2,n}$ and $\hat{\mu}_{n} = \mu_{n} + z_{3,n}$, where $z_{i,n}\sim\mathbf{\mathcal{N}}\left(0, \sigma_{i,n}^{2}\right),i=1,2,3$ denotes the measurement noise.
The measurement noise variances $\sigma_{i,n}^{2}, i=1,2,3$ are modelled as $\sigma_{1,n}^{2} \propto (\gamma_{n}\sin^{2}{\phi_{n}})^{-1}$ and $\sigma_{i,n}^{2} \propto \gamma_{n}^{-1}, i=2,3$, respectively, where $\gamma_{n} = \gamma_{\text{r}}/d_{n}^{2}$ denotes the sensing signal-to-noise ratio \cite{JunkunYan2015TSP}. 
The coefficient $\gamma_{\text{r}}$ is defined as $\gamma_{\text{r}}\triangleq N_{\text{t}}N_{\text{r}}P_{\text{A}}N_{\text{sym}}\beta_{\text{r}}/\sigma^{2}$, where $N_{\text{sym}}$ denotes the matched filtering gain.

\subsection{Target Tracking Model}

We assume that the target movement follows the classic constant-velocity model \cite{JunkunYan2015TSP}. 
In particular, the relative motion state at the $n$th time slot is evolved from the counterpart at the $(n-1)$th time slot as $\mathbf{x}_{n} = \mathbf{G}\mathbf{x}_{n-1} - \mathbf{u}_{\text{A},n} + \mathbf{z}_{\text{p},n}$, where $\mathbf{G}$ denotes the transition matrix, $\mathbf{u}_{\text{A},n}=[(v_{\text{A},n} - v_{\text{A},n-1})\Delta T, (v_{\text{A},n} - v_{\text{A},n-1})]^{T}$ denotes the UAV motion state increment at the $n$th time slot, $v_{\text{A},n}$ denotes the UAV velocity at the $n$th time slot, and $\mathbf{z}_{\text{p},n} \sim \mathcal{N}(\mathbf{0},\mathbf{Q}_{\text{p}})$ denotes the Gaussian process noise with zero mean and the covariance matrix $\mathbf{Q}_{\text{p}}$. 
According to the constant-velocity model, the expressions of $\mathbf{G}$ and $\mathbf{Q}_{\text{p}}$ can be specified as 
\vspace{-1.5mm}
\begin{equation}
	\mathbf{G} = \begin{bmatrix}
        1 & \Delta T  \\
        0 & 1 
    \end{bmatrix}, 
    \mathbf{Q}_{s} = \begin{bmatrix}
        \frac{1}{3}\Delta T^{3} & \frac{1}{2}\Delta T^{2}  \\
        \frac{1}{2}\Delta T^{2} & \Delta T 
    \end{bmatrix}\tilde{q},
    \vspace{-1.5mm}
\end{equation}
respectively, where $\tilde{q}$ denotes the process noise intensity \cite{JunkunYan2015TSP}. 

To track the target with time-varying motion state, online relative motion state estimation is performed based on the extended Kalman filtering (EKF) method \cite{YFJ2024CL}.   
Under the EKF framework, the predicted state variables $\breve{\mathbf{x}}_{n}=[\breve{x}_{n},\breve{v}_{n}]^{T}$ can be obtained as 
\vspace{-1.5mm}
\begin{equation}
    \breve{\mathbf{x}}_{n} = \mathbf{G}\hat{\mathbf{x}}_{n-1} - \mathbf{u}_{\text{A},n},  \label{formu::brevexn}
    \vspace{-1.5mm}
\end{equation}
where $\hat{\mathbf{x}}_{n-1}$ denotes the estimated state variables at the $(n-1)$th time slot. 
Then, the measured results $\mathbf{y}_{n} \triangleq [\hat{\phi}_{n}, \hat{\tau}_{n}, \hat{\mu}_{n}]^{T}$ can be compactly expressed as $\mathbf{y}_{n} = \mathbf{h}(\mathbf{x}_{n}) + \mathbf{z}_{\text{m},n}$, where the function $\mathbf{h}(\cdot)$ can be obtained from $\phi_{n}=\arctan{(\frac{H}{x_{n}})}$, $d_{n}=\sqrt{H^{2} + x_{n}^{2}}$ and $\mu_{n}=-\frac{2v_{n}x_{n}}{\lambda\sqrt{x_{n}^{2} + H^2}}$, and $\mathbf{z}_{\text{m},n} = [z_{1,n}, z_{2,n}, z_{3,n}]^{T} \sim \mathcal{N}(\mathbf{0},\mathbf{Q}_{\text{m},n})$ denotes the measurement noise vector with the covariance matrix denoted by $\mathbf{Q}_{\text{m},n} = \text{diag}(\sigma_{1,n}^{2}, \sigma_{2,n}^{2}, \sigma_{3,n}^{2})$.
Next, the state prediction mean square error (MSE) matrix $\mathbf{M}_{\text{p},n}$ and the Kalman gain matrix $\mathbf{K}_{n}$ can calculated by $\mathbf{M}_{\text{p},n}=\mathbf{G}\mathbf{M}_{n-1}\mathbf{G}^{H} + \mathbf{Q}_{\text{p}}$ and $\mathbf{K}_{n} = \mathbf{M}_{\text{p},n}\mathbf{H}_{n}^{H}(\mathbf{Q}_{\text{m},n} + \mathbf{H}_{n}\mathbf{M}_{\text{p},n}\mathbf{H}_{n}^{H} )^{-1}$, respectively, where $\mathbf{M}_{n-1}$ denotes the estimation MSE matrix at the $(n-1)$th time slot and $\mathbf{H}_{n}$ denotes the Jacobian matrix for $\mathbf{h}(\cdot)$ with respect to the predicted state variables specified as
\vspace{-1.5mm}
\begin{equation} 
	\!\!\!\!\mathbf{H}_{n} = \frac{\partial\mathbf{h}}{\partial\mathbf{x}_{n}}\bigg|_{\mathbf{x}_{n}=\breve{\mathbf{x}}_{n}} \!\!\!\!=\!\! \begin{bmatrix}
        \frac{H}{H^{2} + \breve{x}_{n}^{2}} & \frac{\breve{x}_{n}}{\sqrt{H^{2} + \breve{x}_{n}^{2}}} & \frac{-2 \breve{v}_{n}H^{2}}{\lambda \sqrt{(H^{2} + \breve{x}_{n}^{2})^{3}} } \\
         0 & 0 & \frac{-2 \breve{x}_{n}}{\lambda \sqrt{H^{2} + \breve{x}_{n}^{2}} }
    \end{bmatrix}^{T}\!\!\!\!. 
    \vspace{-1.5mm}
\end{equation} 
Utilizing both $\breve{\mathbf{x}}_{n}$ and $\mathbf{y}_{n}$, the state variables at the $n$th time slot can be estimated by $\hat{\mathbf{x}}_{n} = \breve{\mathbf{x}}_{n} + \mathbf{K}_{n}(\mathbf{y}_{n} - \mathbf{h}(\breve{\mathbf{x}}_{n}))$ and the estimation MSE matrix at the $n$th time slot can be given by $\mathbf{M}_{n} = ( \mathbf{H}_{n}^{H}\mathbf{Q}_{\text{m},n}^{-1}\mathbf{H}_{n} + \mathbf{M}_{\text{p},n}^{-1})^{-1}$. 

The accuracy of the estimated state variables at the $n$th time slot can be characterized by the predicted PCRB. 
In particular, the predicted PCRB for the estimated relative position and velocity are given by $\breve{\mathrm{PCRB}}_{\text{x},n} = [\breve{\mathbf{M}_{n}}]_{11}$ and $\breve{\mathrm{PCRB}}_{\text{v},n} = [\breve{\mathbf{M}_{n}}]_{22}$ with $\breve{\mathbf{M}_{n}} = \mathbf{M}_{n} |_{\mathbf{x}_{n} =\breve{\mathbf{x}}_{n} } \approx \mathbf{M}_{n}$, respectively \cite{YFJ2024CL,JunkunYan2015TSP}.
The specific expressions of predicted PCRBs are given by $\breve{\mathrm{PCRB}}_{\text{x},n} = F_{\text{x}}(\breve{\mathbf{x}}_{n})/D(\breve{\mathbf{x}}_{n})$ and $\breve{\mathrm{PCRB}}_{\text{v},n} = F_{\text{v}}(\breve{\mathbf{x}}_{n})/D(\breve{\mathbf{x}}_{n})$, respectively, where $F_{\text{x}}(\breve{\mathbf{x}}_{n})$, $F_{\text{v}}(\breve{\mathbf{x}}_{n})$ and $D(\breve{\mathbf{x}}_{n})$ are detailed in (\ref{formu::PCRB-nume})-(\ref{formu::PCRB-deno}) at the top of the next page.

\begin{figure*}[!t]
    \vspace{-5mm}
    \begin{equation}
        F_{\text{x}}(\breve{\mathbf{x}}_{n}) = \frac{H^{4}\gamma_{\text{r}}}{a_{1}^{2}(H^{2}+\breve{x}_{n}^{2})^{5}} + \frac{\gamma_{\text{r}}\breve{x}_{n}^{2}}{a_{2}^{2}(H^{2}+\breve{x}_{n}^{2})^{3}} + \frac{4H^{4}\gamma_{\text{r}}\breve{v}_{n}^{2}}{a_{3}^{2}\lambda^{2}(H^{2}+\breve{x}_{n}^{2})^{5}} + r_{n,11}, \ \ 
        F_{\text{v}}(\breve{\mathbf{x}}_{n}) = \frac{4\gamma_{\text{r}}\breve{x}_{n}^{2}}{a_{3}^{2}\lambda^{2}(H^{2}+\breve{x}_{n}^{2})^{3}} + r_{n,22}, \label{formu::PCRB-nume}
    \end{equation}
    \begin{equation}
        \!\!\!\!\!\!\!\!\!\!D(\breve{\mathbf{x}}_{n}) = F_{\text{x}}(\breve{\mathbf{x}}_{n}) F_{\text{v}}(\breve{\mathbf{x}}_{n}) - \left(r_{n,12} + \frac{4 H^{2} \gamma_{\text{r}} \breve{v}_{n} \breve{x}_{n} }{a_{3}^{2}\lambda^{2}(H^{2}+\breve{x}_{n}^{2})^{4}}\right)\left(r_{n,21} + \frac{4 H^{2} \gamma_{\text{r}} \breve{v}_{n} \breve{x}_{n} }{a_{3}^{2}\lambda^{2}(H^{2}+\breve{x}_{n}^{2})^{4}}\right), \ \ 
        r_{n,ij} = [ \mathbf{M}_{\text{p},n}^{-1} ]_{ij}, 
        \label{formu::PCRB-deno}
    \end{equation}
    \begin{equation}
        \!\!\!\!\!\!\!\!\!\!\!\!\!\!\!\!\!\!\!\!\!\!\!\!\!\!\!\!\!\!\!\!\!\!\!\!\!\!\!\!\!\!\!\!\!\!\!\!\!\!\!\!\!\!\!\!\!\!\!\!\!\!P( v_{\text{A},n} ) = P_{0}\left( 1 + 3 v_{\text{A},n}^{2}/U_{\text{tip}}^{2}\right) 
        + P_{\rm{i}}\left( \left( 1 + v_{\text{A},n}^{4}/4 v_{\rm{h}}^{4}\right)^{1/2} - v_{\text{A},n}^{2}/2 v_{\rm{h}}^{2} \right)^{1/2} 
        + \frac{1}{2} \chi |v_{\text{A},n}|^{3}, \label{formu::UAV-propulsion}
        \vspace{-2mm}
    \end{equation}
    \rule{18.2cm}{0.5pt}
    \vspace{-10mm}
\end{figure*}

\vspace{-2mm}
\subsection{Problem Formulation} 

In this paper, we consider online UAV trajectory optimization for target tracking. 
At each time slot, the predicted state variables are optimized to minimize the weighted sum of predicted PCRBs. 
The optimization problem is formulated as
\vspace{-5.5mm}
\begin{align}
    (\mathrm{P1.n}): \ &\min_{ \breve{\mathbf{x}}_{n} } \ \ \alpha\breve{\mathrm{PCRB}}_{\text{x},n} + (1-\alpha)\breve{\mathrm{PCRB}}_{\text{v},n} \label{opt-obj}\\
    \text{s.t.} \ 
    & \sum\limits_{n=1}^{N} E_{n} \leq E_{\text{tot}}, \forall n \in \{1,2,...,N\}, \tag{\ref{opt-obj}{a}} \label{opt-1-a} \\
    & \left| \eta_{n-1} - \breve{x}_{n} \right| \leq v_{\text{A,max}}\Delta T, \tag{\ref{opt-obj}{b}} \label{opt-1-b} \\
    & \breve{x}_{n} - \Delta T\breve{v}_{n} - \hat{x}_{n-1} = 0, \tag{\ref{opt-obj}{c}} \label{opt-1-c} \\
    & x_{\text{A},0} = x_{\text{I}}, \ \ x_{\text{A},N} = x_{\text{F}}, \tag{\ref{opt-obj}{d}} \label{opt-1-d}
    \vspace{-1.5mm}
\end{align}
where $\alpha\in [0,1]$ denotes the regularized weighting factor,\footnote{In practice, the weighting factor can be designed to balance the position and velocity estimation errors according to specific sensing requirements.}
$E_{n} = P( v_{\text{A},n} )\Delta T$ denotes the UAV propulsion energy consumption at the $n$th time slot, $E_{\text{tot}}$ denotes the total propulsion energy consumption budget, the variable $\eta_{n-1}$ is defined as $\eta_{n-1}\triangleq\hat{x}_{n-1} + \hat{v}_{n-1}\Delta T + v_{\text{A},n-1}\Delta T$, $v_{\text{A,max}}$ denotes the UAV maximum velocity and $x_{\text{A},n}$ denotes the UAV position at the $n$th time slot. 
$P( v_{\text{A},n} )$ denotes the UAV propulsion power given by (\ref{formu::UAV-propulsion}), where $P_{0}$, $P_{\rm{i}}$, $U_{\text{tip}}$, $v_{\rm{h}}$ and $\chi$ are related constant parameters specified in \cite{ZY-EM-Rotary-2019-TWC}. 
In (P1.n), (\ref{opt-1-a}) represents the total UAV propulsion energy consumption constraint during the whole flight. 
(\ref{opt-1-b}), (\ref{opt-1-c}) and (\ref{opt-1-d}) represents the UAV maximum velocity constraint, the predicted state variables coupling constraint and the UAV initial-final location constraint, respectively. 
Note that optimizing the predicted state variables is equivalent to the UAV trajectory design due to (\ref{formu::brevexn}) and $x_{\text{A},n} = x_{\text{A},n-1} + v_{\text{A},n}\Delta T$. 

(P1.n) is a non-convex optimization problem due to the non-convex objective function and the constraint (\ref{opt-1-a}). 
Moreover, note that (P1.n) represents an online UAV trajectory optimization problem because, at the $n$th time slot, both the UAV trajectory and propulsion energy consumption at the following time slots have been undetermined. 
However, constraints (\ref{opt-1-a}) and (\ref{opt-1-d}) must be considered when optimizing the UAV trajectory for each time slot. 
Otherwise, the sequentially generated UAV trajectory may not be feasible for both constraints. 
For example, if the propulsion energy is overused at the current time slot, the left propulsion energy may be insufficient for the UAV to reach the final location. 
Thus, it is non-trivial to solve (P1.n) while ensuring the feasibility of the obtained solution.

\section{Proposed Approach} 

In this section, an online UAV trajectory optimization approach is proposed to solve (P1.n). 
Note that, at the $n$th time slot, the undetermined UAV trajectory can be divided into the trajectory to be optimized at the current time slot and the trajectory at future time slots, denoted by $x_{\text{A},n}$ and $\{ x_{\text{A},l} \}, l = n+1,...,N$, respectively. 
If the undetermined UAV trajectory is specified, then the feasibility of the solution to (P1.n) can be judged at the $n$th time slot. 
As a result, the idea of our approach is to design $x_{\text{A},n}$ and $\{ x_{\text{A},l} \}$ by solving two subproblems, named as the \emph{candidate trajectory} and the \emph{energy-aware backup trajectory} optimization problem, respectively.   
The subproblems and the procedures of our proposed approach are specified as follows.  

\subsubsection{Initialization}
We first define a variable $E_{\text{c},n} \triangleq \sum_{m=1}^{n-1}E_{m}$ as the consumed UAV propulsion energy until the $(n-1)$th time slot. 
Particularly, $E_{\text{c},1} = 0$ and $n=1$ is initialized at the first time slot.

\subsubsection{Candidate trajectory optimization} 
At the $n$th time slot, the candidate trajectory optimization problem is formulated as 
\vspace{-5.5mm}
\begin{align}
    (\mathrm{P2.n}): \ &\min_{ \breve{\mathbf{x}}_{n} } \ \ \alpha\breve{\mathrm{PCRB}}_{\text{x},n} + (1-\alpha)\breve{\mathrm{PCRB}}_{\text{v},n} \label{opt-obj-P2} \\ 
    \text{s.t.} \ &\text{(\ref{opt-1-b}),(\ref{opt-1-c})}, \notag \\
    &| \breve{x}_{n} - \omega_{n} | \!\leq \! (N - n) v_{\text{A,max}} \Delta T, \tag{\ref{opt-obj-P2}{a}} \label{opt-obj-P2-a}
    \vspace{-1.5mm}
\end{align}
with $\omega_{n} = x_{\text{A},n-1} + \eta_{n-1} - x_{\rm{F}}$.
In (P2.n), the constraint (\ref{opt-obj-P2-a}) is derived from $|x_{\text{F}} - x_{\text{A},n}| \!\leq \! (N - n) v_{\text{A,max}} \Delta T$, representing that given the UAV position at the $n$th time slot, the UAV can arrive at the final location with its maximum velocity within the following $N-n$ time slots. 
In this way, the solution to (P2.n) is sure to satisfy the constraint (\ref{opt-1-d}) in (P1.n). 

Our proposed solution to (P2.n) is as follows. 
(P2.n) is a non-convex fractional programming problem and can be addressed via Dinkelbach's transform \cite{KaimingShen2018TSP}. 
Specifically, we first reformulate the numerator and denominator of the objective function of (P2.n) as $A(\breve{\mathbf{x}}_{n}) = D(\breve{\mathbf{x}}_{n}) (H^{2} + \breve{x}_{n}^{2})^{8}$ and $B(\breve{\mathbf{x}}_{n}) = (\alpha F_{\text{x}}(\breve{\mathbf{x}}_{n}) + (1-\alpha)F_{\text{v}}(\breve{\mathbf{x}}_{n})) (H^{2} + \breve{x}_{n}^{2})^{8}$, respectively. 
Then, (P2.n) can be formulated as
\vspace{-1.5mm}
\begin{equation}
    (\mathrm{P2.n'}): \ \min_{ \breve{\mathbf{x}}_{n} } \ \ C(\breve{\mathbf{x}}_{n}) \ \ 
    \text{s.t.} \ \ \text{(\ref{opt-1-b}),(\ref{opt-1-c}),(\ref{opt-obj-P2-a})}, \notag
    \vspace{-1.5mm}
\end{equation}
where $C(\breve{\mathbf{x}}_{n}) \triangleq - A(\breve{\mathbf{x}}_{n}) + \zeta B(\breve{\mathbf{x}}_{n})$, and the auxiliary variable $\zeta$ is updated by $\zeta_{k+1} = A(\breve{\mathbf{x}}_{n,k}^{*})/B(\breve{\mathbf{x}}_{n,k}^{*})$ with $\breve{\mathbf{x}}_{n,k}^{*}$ denoting the solution to (P2.n$'$) in the $k$th iteration. 

(P2.n$'$) is a polynomial program, and its global optimal solution can be obtained via Lasserre hierarchy \cite{Lasserre2002}.  
To be specific, we substitute $(\breve{x}_{n} - \hat{x}_{n-1})/\Delta T$ into $\breve{v}_{n}$ due to the constraint (\ref{opt-1-c}). 
Then, a slack variable vector $\mathbf{t}\in\mathbb{R}^{17}$ is introduced and defined as $\mathbf{t} \triangleq [t_{1}, t_{2}, ..., t_{17}]^{T}$ with $t_{q} = \breve{x}_{n}^{q}, q = 1,2,...,17$. 
Given $\mathbf{t}$, the objective function of (P2.n$'$) can be reformulated as $C(\breve{\mathbf{x}}_{n}) = \mathbf{c}^{T}\mathbf{t}$, where $\mathbf{c} = [c_{1}, c_{2}, ..., c_{17}]^{T} \in \mathbb{R}^{17}$ denotes the monomial coefficient with $c_{17} = 0$ since the maximum degree of $\breve{x}_{n}$ in $C(\breve{\mathbf{x}}_{n})$ is $16$.
Afterward, we define two Hankel matrices as 
\vspace{-1.5mm}
\begin{equation}
    \!\!\mathbf{L}(\mathbf{t})\! \triangleq \!\!
    \begin{bmatrix}
        1 & t_{1} & ... & t_{8} \\
        t_{1} & t_{2} & ... & t_{9} \\
        ... & ... & ... & ... \\
        t_{8} & t_{9} & ... & t_{16} 
    \end{bmatrix}\!\!, 
    \mathbf{M}(\mathbf{t})\! \triangleq \!\!
    \begin{bmatrix}
        t_{1} & t_{2} & ... & t_{9} \\
        t_{2} & t_{3} & ... & t_{10} \\
        ... & ... & ... & ... \\
        t_{9} & t_{10} & ... & t_{17} 
    \end{bmatrix}\!\!, 
    \vspace{-1.5mm}
\end{equation}
respectively. 
As such, (P2.n$'$) can be reformulated as 
\vspace{-1.5mm}
\begin{equation}
    (\mathrm{P2.n''}): \ \min_{ \mathbf{t} } \ \ \mathbf{c}^{T}\mathbf{t} \ \ 
    \text{s.t.} \ \ \overline{t} \ \mathbf{L}(\mathbf{t}) \succeq \mathbf{M}(\mathbf{t}) \succeq \underline{t} \ \mathbf{L}(\mathbf{t}), \label{opt-obj-P22}
    \vspace{-1.5mm}
\end{equation}
where the expressions of $\underline{t}$ and $\overline{t}$ are shown in (\ref{formu::tminmax}). 
The constraint in (\ref{opt-obj-P22}) is due to $\underline{t} \leq \breve{x}_{n} \leq \overline{t}$ \cite[Proposition 2.1]{Lasserre2002}, which is the equivalent incorporation of (\ref{opt-1-b}) and (\ref{opt-obj-P2-a}).
(P2.n$''$) is a semidefinite program (SDP) that can be optimally solved by CVX tools. 
Moreover, a sufficient condition for the obtained solution to be globally optimal can be given by $A(\breve{\mathbf{x}}_{n})$ and $B(\breve{\mathbf{x}}_{n})$ being concave and convex, respectively, due to the property of Dinkelbach's transform \cite{KaimingShen2018TSP}. 

By solving (P2.n), a candidate UAV trajectory denoted by $x_{\text{A},n}^{*} = x_{\text{A},n-1} + \eta_{n-1} - \breve{x}_{n}^{*}$ is obtained for minimizing the weighted sum-predicted PCRBs at the $n$th time slot without the total energy consumption constraint (\ref{opt-1-a}), where $\breve{\mathbf{x}}_{n}^{*} = [\breve{x}_{n}^{*}, \breve{v}_{n}^{*}]^{T}$ denotes the solution to (P2.n). 
Whether this candidate UAV trajectory $x_{\text{A},n}^{*}$ is feasible for (P1.n) will be addressed after solving the following optimization problem. 

\begin{figure*}[!t]
    \vspace{-5mm}
    \begin{equation}
        \underline{t} = \max{ \{ \eta_{n-1} - v_{\rm{A,max}}\Delta T, \omega_{n} - (N-n)v_{\rm{A,max}}\Delta T \} }, \overline{t} = \min{ \{ \eta_{n-1} + v_{\rm{A,max}}\Delta T, \omega_{n} + (N-n)v_{\rm{A,max}}\Delta T \} }, \label{formu::tminmax}
        \vspace{-3mm}
    \end{equation}
    \rule{18.2cm}{0.5pt}
    \vspace{-10mm}
\end{figure*}

\subsubsection{Energy-aware backup trajectory optimization}

Given the solution to (P2.n), the energy-aware backup trajectory optimization problem can be formulated as
\vspace{-1.5mm}
\begin{align}
    (\mathrm{P3.n}): \ &\min_{ \{ v_{\text{A},l} \} } \ \ \sum\limits_{l=n+1}^{N} P(v_{\text{A},l})\Delta T \label{opt-obj-P3} \\
    \text{s.t.} \ 
    & |v_{\text{A},l}| \leq v_{\text{A,max}}, \forall l, \tag{\ref{opt-obj-P3}{a}} \label{opt-obj-P3-a} \\
    & x_{\text{A},l-1} = x_{\text{A},n}^{*}, \ \ x_{\text{A},N} = x_{\text{F}} \tag{\ref{opt-obj-P3}{b}} \label{opt-obj-P3-b}.
    \vspace{-1.5mm}
\end{align}
(P3.n) can be solved offline since it is irrelevant to the predicted PCRBs, which are calculated online following the EKF procedures. 
To address the non-convex objective function of (P3.n), slack variables denoted by $\{\xi_{l}\}$ are introduced such that $\xi_{l} \geq ( (1 +  v_{\text{A},l}^{4}/(4 v_{\rm{h}}^{4}) )^{1/2} - v_{\text{A},l}^{2}/(2 v_{\rm{h}}^{2}) )^{1/2} \geq 0 $. 
As a result, the constraint $\xi_{l}^{-2} \leq \xi_{l}^{2} + v_{\text{A},l}^{2}/v_{\rm{h}}^{2}$ should be satisfied. 
Note that $\xi_{l}^{2} + v_{\text{A},l}^{2}/v_{\rm{h}}^{2}$ is convex regarding to both $\xi_{l}$ and $v_{\text{A},l}$, and lower bounded by its first-order Taylor expansion at a given point $\{\xi_{l,r}, v_{\text{A},l,r}\}$ in the $r$th iteration.
Thus, we apply the SCA technique to approximate this constraint by 
\vspace{-1.5mm}
\begin{equation}
    \!\!\!\xi_{l}^{-2} \!\leq\! 2\xi_{l,r}(\xi_{l} - \xi_{l,r}) + \xi_{l,r}^{2} + 2(v_{\text{A},l} - v_{\text{A},l,r})/v_{\rm{h}}^{2} + v_{\text{A},l,r}^{2}/v_{\rm{h}}^{2}. \label{formu::P3-SCA}
    \vspace{-1.5mm}
\end{equation}
Given (\ref{formu::P3-SCA}), the solution to (P3.n) can be obtained by iteratively solving the following optimization problem formulated as 
\vspace{-1.5mm}
\begin{equation}
    (\mathrm{P3.n'}): \ \min_{ \{ v_{\text{A},l} \}, \{ \xi_{l} \} } \ \ \tilde{P} \label{opt-obj-P31}
\end{equation}
\vspace{-4mm}
\begin{equation}
    \ \ \ \text{s.t.} \
    \xi_{l} \geq 0, \forall l, \tag{\ref{opt-obj-P31}{a}} \label{opt-obj-P31-a}
    \vspace{-1.5mm}
\end{equation}
\vspace{-5mm}
\begin{equation}
    \ \ \ \ \ \ \ \ \ \ \ \ \text{(\ref{formu::P3-SCA}),(\ref{opt-obj-P3-a}),(\ref{opt-obj-P3-b})}, \notag 
    \vspace{-1.5mm}
\end{equation}
with $\tilde{P} = \sum_{l=n+1}^{N} P_{0} ( 1 + 3 v_{\text{A},l}^{2}/U_{\text{tip}}^{2} ) + P_{i}\xi_{l} + \chi|v_{\text{A},l}|^{3}/2$. 
(P3.n$'$) is a convex optimization problem and can be solved by CVX tools.
Given the solutions to (P3.n) denoted by $\{v_{\text{A},l}^{*}\}$, an energy-aware backup trajectory is obtained by $\{ x_{\text{A},l}^{*} \}$ with $\{ x_{\text{A},l}^{*} \}$ with $x_{\text{A},l}^{*} = x_{\text{A},l-1}^{*} + v_{\text{A},l}^{*}, l=n+1,...,N$. 
The reason for such design is that it is easier for the candidate trajectory $x_{\text{A},n}^{*}$ to be feasible if the UAV consumes as least propulsion energy as possible at the following time slots.

\subsubsection{Feasibility verification and output}
After solving (P2.n) and (P3.n), the feasibility of the candidate trajectory $x_{\text{A},n}^{*}$ can now be verified. 
Specifically, let $E_{\text{b},n} = \sum_{l=n+1}^{N}P(v_{\text{A},l}^{*})\Delta T$ denote the propulsion energy consumption for the energy-aware backup trajectory. 
If 
\vspace{-1.5mm}
\begin{equation}
 E_{\text{c},n} + P\left(\frac{x_{\text{A},n}^{*} - x_{\text{A},n-1}}{\Delta T}\right)\Delta T + E_{\text{b},n} \leq E_{\text{tot}}, \label{formu::EtotJudge}
 \vspace{-1.5mm}
\end{equation}
holds, then the candidate trajectory $x_{\text{A},n}^{*}$ satisfies the constraint (\ref{opt-1-a}) and, therefore, can be output as the solution to (P1.n) with its feasibility guaranteed. 
Otherwise, the candidate trajectory is infeasible because it leads to the total propulsion energy consumption exceeding the budget $E_{\text{tot}}$. 
In this case, to ensure the feasibility of (P1.n), we resort to output the energy-aware backup trajectory obtained at the former time slot, i.e., $\{x_{\text{A},\kappa}^{*}\}, \kappa = n, ..., N$ as the designed UAV trajectory at the rest time slots.\footnote{At the first time slot, if (\ref{formu::EtotJudge}) is not satisfied, the UAV trajectory can be designed as the solution to (P3.n) with $n=0$ and (\ref{opt-obj-P3-b}) replaced by (\ref{opt-1-d}).}
Note that the objective function in (\ref{opt-obj-P31}) is an upper bound of the objective function in (\ref{opt-obj-P3}) due to the slack operation. 
Thus, (\ref{formu::EtotJudge}) is a sufficient condition for the optimized UAV trajectory to be feasible. 

\begin{figure}[!t]
    \centering
    \includegraphics[width=0.46\textwidth]{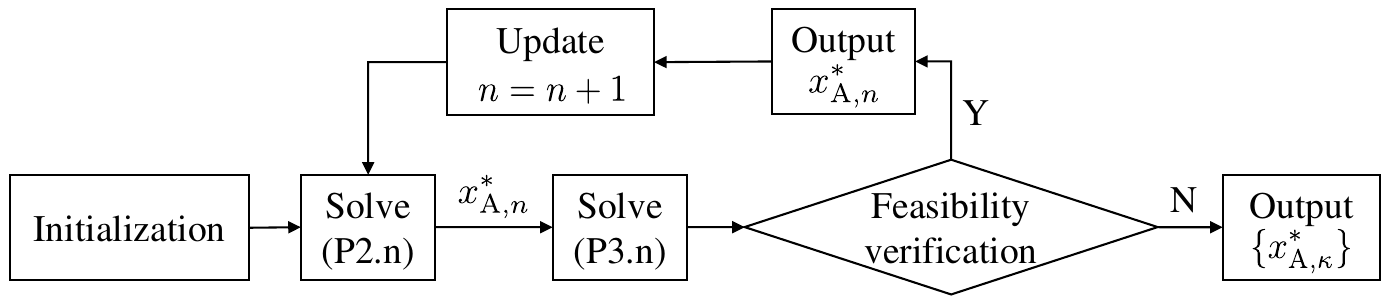}
    \vspace{-1mm}
    \caption{The proposed approach procedures.}
    \label{fig::proposed-online}
    \vspace{-5mm}
\end{figure}

The aforementioned procedures are summarized in Fig.\ref{fig::proposed-online}. 
At each time slot, the UAV trajectory is optimized by solving (P2.n) for sensing performance maximization, and its feasibility of satisfying (\ref{opt-1-a}) is verified by solving (P3.n) and further judging whether (\ref{formu::EtotJudge}) holds. 
Note that $x_{\text{A},n}^{*}$ is regardless of the propulsion energy consumption, while $\{x_{\text{A},\kappa}^{*}\}$ minimizes the propulsion energy to be consumed.
Therefore, our approach frees the UAV trajectory optimization from the total energy consumption constraint when there is enough energy left in real time, which offers excellent flexibility for sensing performance maximization.

\section{Numerical Results}

This section provides numerical results to illustrate the effectiveness of our proposed UAV trajectory optimization approach. 
Unless specified otherwise, the numerical values of key system parameters are set as follows: $P_{0} = 79.8563$ W, $P_{i} = 88.6279$ W, $U_{\text{tip}} = 120$ m/s, $v_{\rm{h}} = 4.03$ m/s, $\chi = 0.0185$ kg$\cdot \rm{m}^{2}$ \cite{ZY-EM-Rotary-2019-TWC}, $P_{\text{A}} = 20$ dBm \cite{QiaoyanPeng2024-TCOM,HonghaoWang2024WCL,MengHua2024TWC,YapengZhao2024arxiv,YifanJiang2021AC,YifanJiang2020ICCC}, $N_{\text{sym}} = 10^{4}$, $\Delta T = 0.2$ s, $\lambda = 0.01$ m, $\sigma^{2} = -80$ dBm, $\tilde{q} = 1$, $\varepsilon = 100 \ \text{m}^{2}$, $N_{\text{t}}=N_{\text{r}}=16$, $a_{1} = 0.1$, $a_{2} = 10$, $a_{3} = 2000$, $\alpha = 0.5$, and $H=50$ m \cite{YFJ2024CL}. 
Our proposed scheme is compared with a benchmark: at the $n$th time slot, if there is enough propulsion energy for direct flight to the final location at the following time slots, represented by $E_{\rm{tot}} - E_{\text{c},n} > P(v_{\text{A,max}})\Delta T + (N-n)P(v_{\text{df}})\Delta T$ with $v_{\text{df}} = (x_{\text{F}} - x_{\text{A},n})/(N-n)\Delta T$, then the UAV trajectory is obtained by solving (P1.n) with the constraint (\ref{opt-1-a}) replaced with $E_{n} \leq E_{\rm{tot}} - E_{\text{c},n}$.
Otherwise, the UAV trajectory is designed as $v_{\text{A},n}=(x_{\text{F}} - x_{\text{A},n-1})/(N-n+1)\Delta T$.

\begin{figure*}[htbp]
	\centering
	\vspace{-4mm}
	\subfigure[UAV and target trajectories.]{
		\label{simufig1}
		\vspace{-1mm}
		\includegraphics[width=5.5cm]{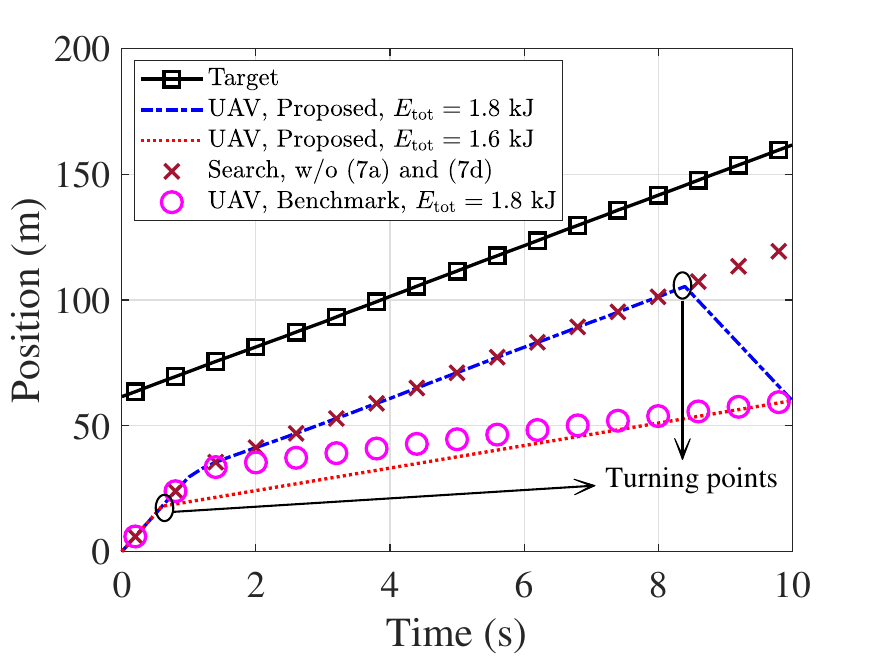}
	}
	\subfigure[Energy consumption and sensing performance.]{
		\label{simufig2}
        \vspace{-1mm}
		\includegraphics[width=5.5cm]{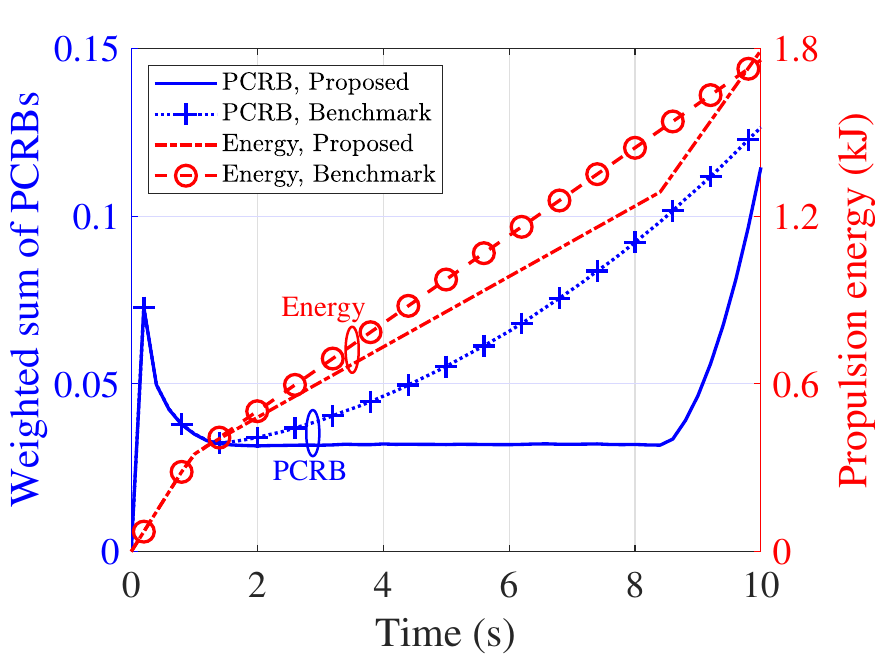}
	}
	\subfigure[Energy consumption and UAV velocity.]{
		\label{simufig3}
        \vspace{-1mm}
		\includegraphics[width=5.5cm]{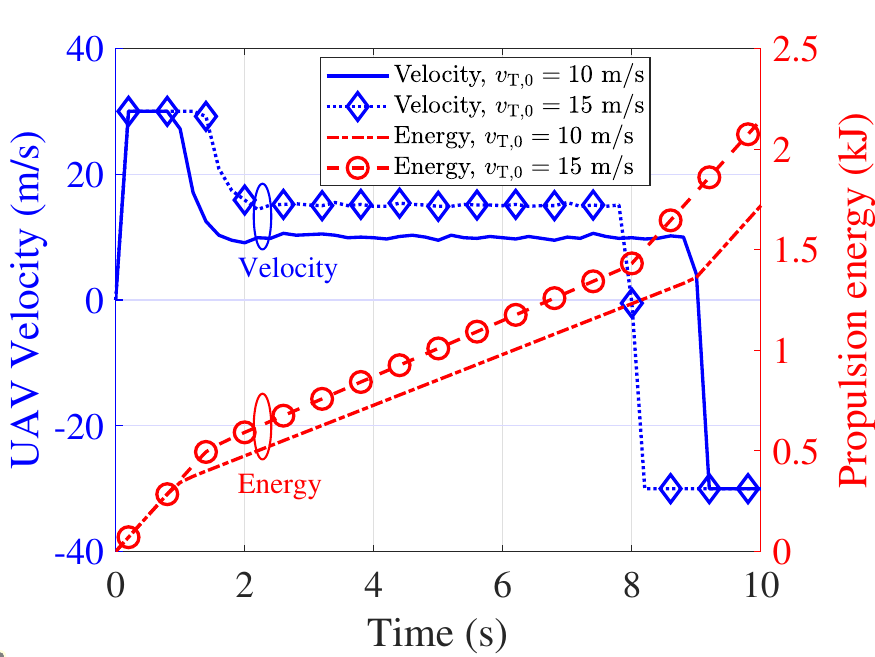}
	}
    \vspace{-2mm}
	\caption {Comparisons between the UAV trajectory, the consumed propulsion energy, and the sensing performance under different cases. }
	\label{figureSimu}
    \vspace{-4mm}
\end{figure*}

In Fig.\ref{simufig1}, we show the target and UAV trajectories under the benchmark and our proposed optimization approach in one trial, with $x_{\rm{I}} = 0$ m, $x_{\rm{F}} = 60$ m, and the results picked every $0.6$ s. 
The target average velocity is given by $v_{\rm{T},0}=10$ m/s. 
As shown in Fig.\ref{simufig1}, our proposed approach results in the turning points of UAV trajectory under both cases with $E_{\rm{tot}} = 1.8$ kJ and $E_{\rm{tot}} = 1.6$ kJ. 
Particularly, the UAV trajectory ahead of the turning point matches the results obtained by solving (P1.n) via the exhaustive search method without the constraints (\ref{opt-1-a}) and (\ref{opt-1-d}), which verifies the effectiveness of the proposed solution to (P2.n). 
After the turning points, the UAV directly flies to the final location due to the constraint (\ref{opt-1-d}). 
In addition, under the case with $E_{\rm{tot}} = 1.6$ kJ, the turning point of the UAV trajectory appears earlier than the case with $E_{\rm{tot}} = 1.8$ kJ due to a smaller energy budget. 
Moreover, compared to the benchmark, our proposed approach allows a much longer duration for the UAV trajectory to be optimized for maximizing the sensing performance under the case with $E_{\rm{tot}} = 1.8$ kJ, rendering it superior to the benchmark. 

Fig.\ref{simufig2} illustrates the propulsion energy consumption and the weighted sum of actual PCRBs achieved by the benchmark and our proposed approach under the cases with $E_{\rm{tot}} = 1.8$ kJ shown in Fig.\ref{simufig1}. 
The actual PCRBs are obtained by the diagonal elements of the estimation MSE matrix at the $n$th time slot $\mathbf{M}_{n}$. 
It is found that the total propulsion energy consumption under our proposed approach is smaller than $1.8$ kJ, which validates the effectiveness of the solution to (P3.n) and the whole proposed UAV trajectory optimization approach. 
Additionally, compared to the benchmark, our proposed approach leads to less energy consumption ahead of the UAV turning points while utilizing more energy in total given the same energy budget $E_{\rm{tot}}$. 
Besides, our proposed approach leads to a significantly lower weighted sum of actual PCRBs than the benchmark. 
These results show that our proposed approach achieves a substantial sensing performance improvement and a more flexible and sufficient utilization of propulsion energy than the benchmark. 

In Fig.\ref{simufig3}, the propulsion energy consumption and the UAV velocity are compared when the target velocity is set as $10$ m/s and $15$ m/s, respectively. 
It is demonstrated that in both cases, the UAV flies at its maximum velocity either when approaching the target at the beginning or when flying towards the final location at the final part of the flight.
However, the UAV velocity remains approximately the same as the target average velocity for the rest of the flight.
Furthermore, the propulsion energy is consumed slower when tracking the target under the $v_{\rm{T},0} = 10$ m/s case because the UAV velocity is closer than $15$ m/s to the maximum-endurance velocity, which is approximately $10.21$ m/s \cite{ZY-EM-Rotary-2019-TWC}.
This phenomenon shows that, apart from the initial-final location constraint, the target velocity fundamentally affects the UAV energy consumption. 

\section{Conclusion}

In this work, a novel online UAV trajectory optimization approach was proposed for ensuring the feasibility of satisfying both the total propulsion energy constraint and the initial-final location constraint.
Our approach achieves this goal by solving two subproblems.
One is for sensing performance maximization, and the other is for feasibility verification. 
Then, two efficient solutions were proposed to solve the subproblems.  
Simulation results verified the effectiveness and superiority of our proposed approach to the benchmark.
Applying our proposed approach in multi-UAV collaborated sensing scenarios is worthwhile for future studies.

\bibliographystyle{IEEEtran}
\bibliography{IEEEabrv,ref}

\end{document}